# Few-layer antimonene electrical properties

Pablo Ares,[1]* Sahar Pakdel,[1,2] Irene Palacio,[3] Wendel S. Paz,[4] Maedeh Rassekh,[1,5] David Rodríguez-San Miguel,[6] Lucía Aballe,[7] Michael Foerster,[7] Nerea Ruiz del Árbol,[3] José Ángel Martín-Gago,[3] Félix Zamora,[6] Julio Gómez-Herrero,[1] Juan José Palacios[1]*

[1]Departamento de Física de la Materia Condensada and Condensed Matter Physics Center (IFIMAC), Universidad Autónoma de Madrid, 28049, Madrid, Spain

[2]Department of Physics and Astronomy, Aarhus University, 8000 Aarhus C, Denmark

[3]ESISNA group, Institute of Material Science of Madrid (ICMM-CSIC), C/Sor Juana Inés de la Cruz 3, 28049 Madrid, Spain

[4]Instituto de Física, Universidade Federal do Rio de Janeiro, Caixa Postal 68528, Rio de Janeiro, RJ 21941-972, Brazil. Departamento de Física, Universidade Federal do Espírito Santo, Vitória, ES 29075-910, Brazil

[5]Department of Physics, University of Guilan, 41335-1914 Rasht, Iran

[6]Departamento de Química Inorgánica, Institute for Advanced Research in Chemical Sciences (IAdChem) and Condensed Matter Physics Center (IFIMAC), Universidad Autónoma de Madrid, 28049, Madrid, Spain

[7]ALBA Synchrotron, Carrer de la llum 2-26, Cerdanyola del Vallès, Barcelona 08290, Spain

*E-mail: pablo.ares@uam.es, E-mail: juanjose.palacios@uam.es

**Abstract**

Antimonene -a single layer of antimony atoms- and its few layer forms are among the latest additions to the 2D mono-elemental materials family. Numerous predictions and experimental evidence of its remarkable properties including (opto)electronic, energetic or biomedical, among others, together with its robustness under ambient conditions, have attracted the attention of the scientific community. However, experimental evidence of its electrical properties is still lacking. Here, we characterized the electronic properties of mechanically exfoliated flakes of few-layer (FL) antimonene of different thicknesses (~ 2-40 nm) through photoemission electron microscopy, kelvin probe force microscopy and transport measurements, which allows us to estimate a sheet resistance of ~ 1200 $\Omega$ sq$^{-1}$ and a mobility of ~ 150 cm$^2$V$^{-1}$s$^{-1}$ in ambient conditions, independent of the flake thickness. Alternatively, our theoretical calculations indicate that topologically protected surface states (TPSS) should play a key role in the electronic properties of FL antimonene, which supports our experimental findings. We anticipate our work will trigger further experimental studies on TPSS in FL antimonene thanks to its simple structure and significant stability in ambient environments.

**1. Introduction**

Recent works have shown how to produce antimonene and few-layer (FL) antimonene with a variety of different procedures [1-6], allowing the exploration of its promising properties [1, 7-11], including its application in optoelectronics [12-15]. One of the properties making antimonene unique among other 2D materials is its strong spin–orbit coupling (SOC). Elemental bulk antimony is a topological semimetal on account of an inverted bulk band order [16]. On the other hand, a band gap in the range of 0.76 to 2.28 eV has been predicted for single layer antimonene [17-23], which makes it very attractive for electronics and ultrafast optoelectronic applications. The electronic properties of antimonene change quite drastically from the single layer to the FL regime. Although antimony does not present a bulk gap, its nonzero topological invariant guarantees the presence of topologically protected surface states (TPSS), coexisting with bulk bands at the Fermi energy [24-26]. For a small number of atomic layers, antimonene could already behave as a 3D topological insulator because quantum confinement opens a gap in its bulk bands [27], but, when this occurs, the TPSS on opposite surfaces couple to each other and a gap opens at the Dirac point, partly degrading their topological properties. A minimum of approximately 7-8 layers (~ 3 nm) is needed for a full decoupling of the TPSS on opposite surfaces [25, 28], but bulk bands already cross the Fermi energy at this small thickness [26, 27]. In this sense, FL antimonene is not so different from actual 3D topological insulators where the contribution of bulk bands at the Fermi energy is difficult to eliminate. Angle Resolved Photoemission Spectroscopy (ARPES) [9, 25, 26] measurements and, to a much lesser extent, transport experiments [27, 28] have confirmed this issue in binary compounds such as $Bi_{1-x}Sb_x$, $Bi_2Se_3$ or $Bi_2Te_3$ [26, 29, 30]. Ternary [31] and quaternary [32] compounds have also shown surface contribution in their conduction properties. Since the surface/bulk conductivity ratio is typically a small fraction, it is difficult to isolate and benefit from the properties of the TPSS [33]. In fact, high-throughput numerical searches have shown that, in theory, stoichiometric topological semimetals outnumber topological insulators [34], thus is of much interest to show that topological semimetals still exhibit properties related with their non-trivial topology [35]. Therefore, finding topological semimetals with a large surface-to-bulk conductivity ratio where the exotic properties of TPSS can manifest in a more direct manner and, additionally, with simple structures and high stability, would increase their possibilities of practical use.

In this work, we present a local morphological and electronic study of mechanically exfoliated FL antimonene flakes of thicknesses between ~ 2 and 40 nm (see Methods). Within this thickness range, the band structure fully reveals the decoupling of the top and bottom TPSS and the bulk bands present a finite contribution at the Fermi energy.

## 2. Results

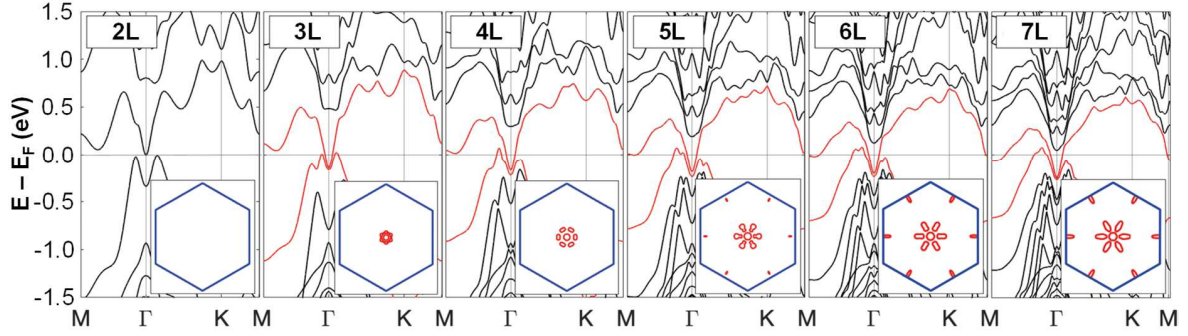

**Figure 1. Band structure of FL antimonene.** Band structure from 2 to 7 layers obtained from DFT calculations (see Methods). The insets show the Fermi surfaces obtained in each case. Red lines correspond to the bands considered for the conductivity calculations. The surface states are located near Γ, and purely bulk pockets near M. Notice that for the latter we observe six half pockets in the first Brillouin zone.

Figure 1 shows DFT calculations (see Methods) of the electronic band structure of FL antimonene in vacuum for different number of layers. The same calculation including adsorbed water molecules on the surface, to partially mimic ambient conditions, does not present significant changes (see Supplementary Figure 1 and Supplementary Figure 2). The insets represent the Fermi surface of each thickness, in the frame of the Brillouin zone (BZ, in blue). The band structure evolution with the number of layers shows two important features. First, the formation of Dirac cones around Γ with the Dirac point about 0.3 eV below the Fermi energy. The Dirac hole band bends upwards to also cross the Fermi energy, giving rise to a peculiar Fermi surface with six hole pockets surrounding Γ. These pockets have a clear surface and helical character near Γ, but this character is partially lost as you move away from Γ and they progressively become bulk states [27]. Second, the appearance of three fully bulk electron pockets crossing the Fermi level close to the M points for over 5 layers. Moving away from Γ towards the M-point (K-point), ignoring the crossing of the bulk electron pockets, one crosses the Fermi level three times (once) as expected for a topological material. The localization properties of these surface states have been studied in Ref. [27].

With this picture in mind, we proceed to the experimental characterization of our exfoliated FL antimonene flakes (see Supplementary Figure 3 for characterization of FL antimonene flakes with different techniques). We start by analyzing flakes with different number of layers and lateral sizes in the range of several microns, deposited on highly doped Si substrates, with low energy and photoemission electron microscopy (LEEM/PEEM), in combination with synchrotron based X-ray photoelectron spectroscopy (XPEEM) (Figure 2). The samples were first cleaned in ultra-high vacuum (UHV) (see Methods), ensuring flakes were free of oxide and contaminants, as demonstrated by the Sb 4d core level spectrum (inset in Figure 2a), which shows the doublet with the 4d 5/2 at a binding energy of around 32 eV. Figure 2a shows a LEEM image of several FL antimonene flakes. XPEEM images and spectra of the same flakes acquired in the maximum of the Sb 4d core level (Supplementary Figure 4) reveal the antimony nature of all the flakes. Figure 2b shows the AFM topography corresponding to the same flakes in Figure 2a, presenting thicknesses from tens of nm down to terraces of just 1.5-2.5 nm (insets at the bottom left and top left corners respectively in Figure 2b). We also carried out the electronic structure characterization by measuring the PEEM signal at the Fermi energy level ($E_F$) (Figure 2c). The direct comparison of the AFM topography and the PEEM signal (see Figure 2d) shows that the photoemitted intensity is mainly constant across the flakes and independent of the thickness in the range ~ 20 to 40 nm, while it is strongly quenched in certain areas marked by white arrows. These areas correspond to the thinnest zones, only 2-5 atomic layers (as shown in the top left inset in Figure 2b), since the minimum height of a monolayer, when measured with AFM, is ~ 1 nm due to molecules trapped between the substrate and the flake [1]. A detailed analysis can be found in Supplementary Figure 5.

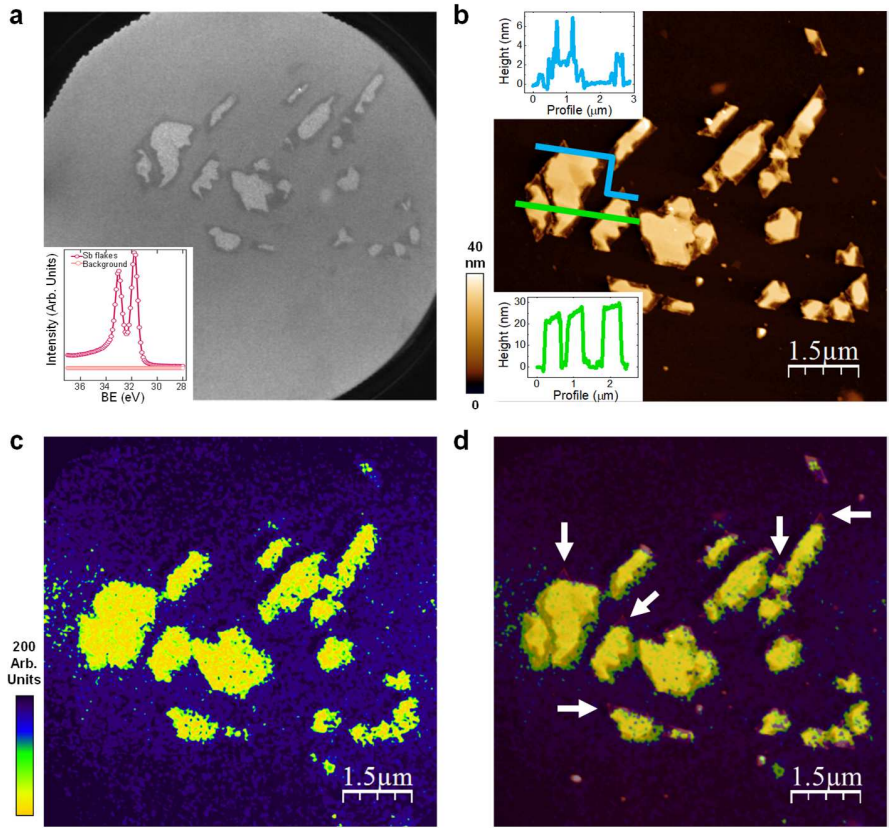

**Figure 2. PEEM/LEEM characterization of FL antimonene flakes of different thicknesses.**
**a**, LEEM image of FL antimonene flakes acquired with an electron energy of 12.4 eV. The image field of view is 10 μm. Inset: The Sb 4d core level micro-spot XPS spectrum taken in the flakes, with a photon energy of 97 eV. **b**, AFM topographical image of the same flakes shown in **a**. The insets correspond to profiles along the lines in the image showing terraces of 1.5-2.5 nm (top) and 20-30 nm (bottom). **c**, PEEM image taken at the Fermi Energy ($E_F$) of the same flakes shown in **a** and **b**, with a photon energy of 97 eV. The contrast is constant for the higher terraces. **d**, Superposition of **b** and **c**. The white arrows are pointing out some of the lowest terraces where no photoemission signal is detected at the $E_F$.

These results are consistent with the DFT calculations shown in Figure 1. For 1-2 layers the material is fully insulating, while for 3-4 layers, corresponding to the majority of the thinnest flakes in Figure 2, the Fermi surface (see Figure 1) is barely developed. For thicker flakes, the integrated contribution to the density of states at the Fermi energy comes mostly from the electron pocket at Γ and the hole pockets around it. Note also that, due to the low electron mean free path at the recorded kinetic energy, the PEEM signal comes from the few topmost 2 or 3 surface atomic layers, around 100 meV below the $E_F$. Hence, as the TPSS have their weight centered there, it is not preposterous to think that this PEEM signal could be associated with the TPSS of the top layer and not from the bulk pockets that extend throughout the flake width.

Next, we carry out an electrical characterization of FL antimonene flakes of different thicknesses. Kelvin probe force microscopy (KPFM) measurements on flakes with thicknesses between ~ 2 to 9 nm (~ 2-3 to 21 layers) [1] (see Supplementary Material section S4) provides information on the Contact Potential Difference (CPD) variation, which is related to the work function change. As in other 2D materials [36], the measured CPD varies with increasing thickness and above a given number of layers, it saturates (see Supplementary Figure 6 and Supplementary Figure 7). We also performed conductive AFM (C-AFM) measurements contacting the flakes with gold nanowire electrodes using the scanning-probe-assisted nanowire circuitry (SPANC) technique [37] and a metal-coated AFM tip as a second mobile electrode on flakes deposited on $SiO_2$. Figure 3a shows FL antimonene flakes of different thicknesses contacted through SPANC. Gold nanowire electrodes are positioned on one end of the flakes and the conductive AFM tip is used to acquire current *vs.* voltage (*IV*) curves at different distances of the gold electrodes (see Supplementary Figure 8) with nm resolution in ambient conditions (see Methods for details on the *IV* curves acquisition procedure). Supplementary Figure 9 shows one of the flakes from Figure 3 imaged more than one year later, showing no noticeable degradation, which accounts for the high stability of antimonene under ambient conditions [1-5].

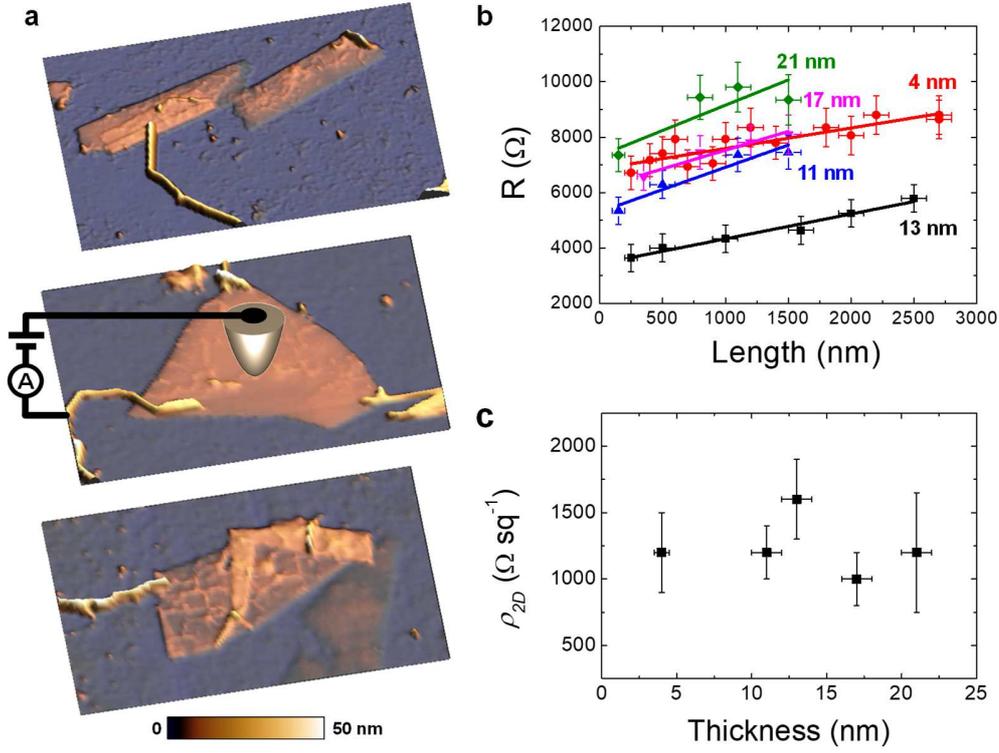

**Figure 3. Electrical transport characterization of FL antimonene of different thicknesses. a**, AFM 3D topographical images of several FL antimonene flakes contacted with gold nanowire electrodes. In the middle panel we have included a schematic of the C-AFM circuit used. Image size: $6.0 \times 3.7$ μm$^2$. **b**, Resistance *vs*. length (*RL*) plot for different thicknesses as indicated in the graph. Solid lines are the best linear fits. The horizontal error bars indicate the uncertainty in the measured distances with the AFM; the vertical error bars indicate the standard deviation in the resistance values from fluctuations in the measured current. **c**, Sheet resistance as a function of the thickness. The horizontal error bars indicate the uncertainty in the measured heights with the AFM; the vertical error bars come from propagating the uncertainties of the slopes of the *RL* plots and the dimensions of the flakes measured with the AFM used to obtain the sheet resistance values.

This local measurement of the conductive properties allows us to obtain resistance *vs.* length (*RL*) plots from different thicknesses (Figure 3b). From the *RL* plots and the geometry of the flakes, we calculate the sheet resistance for the different thicknesses of the FL antimonene flakes (Figure 3c), obtaining robust values regardless of the contact resistance (Supplementary Figure 10). Within the sensitivity of our measurements, we obtain an average value of $\rho_{2D} = 1200 \pm 300$ Ω sq$^{-1}$, *independent of the flake thickness.* This result is remarkable considering that several bulk bands cross the Fermi energy in our thin flakes. For instance, the evolution of the bottom of the second bulk conduction band (located at M) with the number of layers (Supplementary Figure 6) shows that, for 16 layers (approximately 6 nm, which closely corresponds to our thinnest flake), this second band already crosses the Fermi level. Thus, a higher number of bulk bands is expected to be present in the thicker flakes. Even so, the increasing number of bulk bands does not seem to appreciably increase the conductivity, suggesting a dominant contribution of the TPSS.

## 3. Discussion

It is instructive to compare the mentioned results with the case of graphene, where a linear increase of the sheet conductance (the inverse of the sheet resistance) has been observed with the number of layers [38, 39] in contrast with our case, where it remains constant. We compare the sheet resistance ($\rho_{2D}$ = 1200 ± 300 Ω sq$^{-1}$) with that obtained following the same procedure for FL graphene[37], $\rho_{2D\text{-}G}$ = 670 ± 60 Ω sq$^{-1}$. Graphene has 4 Dirac cones, with intervalley scattering typically suppressed by long-range scattering potentials. The Dirac cone of FL antimonene is surrounded by hole pockets which, to a first approximation, altogether give a similar carrier concentration to that of graphene in ambient conditions (from the band structure we estimate $n_{2D}$ ~ 3.5×10$^{13}$ cm$^{-2}$). The helicity of the states in the hole pockets near the Dirac cone are opposite to that of the nearby Dirac states which also prevents inter-pocket back-scattering[19]. Thus, the resistivity of FL antimonene is expected to be comparable but larger than that of graphene, as observed experimentally. Combining the estimation of the carrier concentration from the band structure with the sheet resistance obtained from the transport measurements we can also tentatively estimate the mobility, $\mu$, by $\mu = \frac{\sigma_{2D}}{en_{2D}}$, where $\sigma_{2D}$ is the bidimensional conductance ($\sigma_{2D}$ = 1/$\rho_{2D}$), $e$ the electron charge, and $n_{2D}$ the bidimensional carrier concentration. We obtain a high mobility of $\mu$ ~ 150 cm$^2$V$^{-1}$s$^{-1}$ in ambient conditions. To put this figure in perspective, one can compare it with the mobility of black phosphorus, which is reported to be $\mu_{BP}$ ~ 50 to 1000 cm$^2$V$^{-1}$s$^{-1}$ depending on the thickness [40].

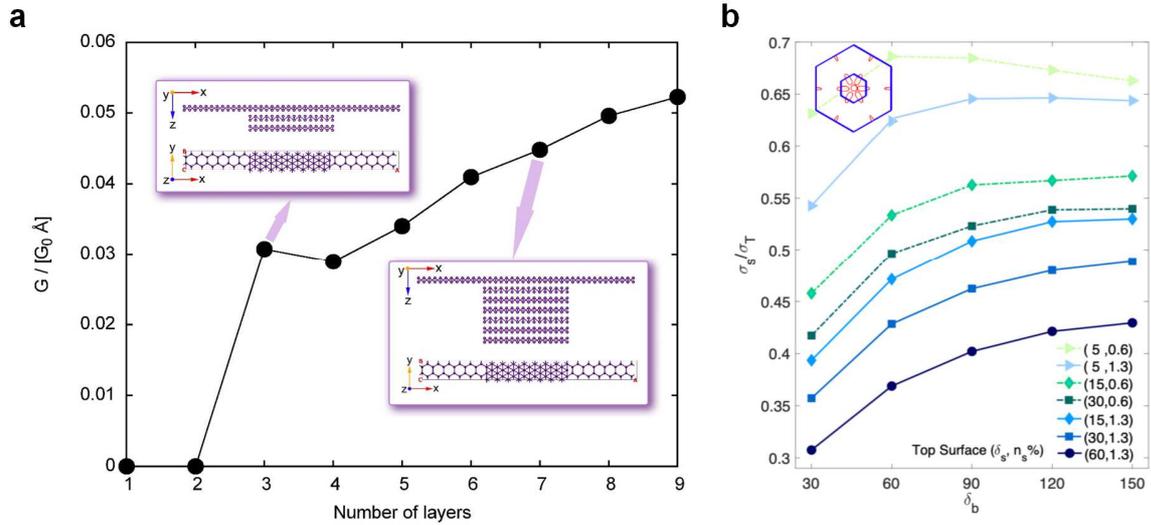

**Figure 4. Transport calculations. a**, Ballistic conductance (per unit length in units of $G_0$) of multilayers contacted on the surface by gated (conducting) antimonene, as sketched in the insets. **b**, Surface to total conductivity ratio ($\sigma_S/\sigma_T$) in the 9L system as a function of the bottom surface disorder strength $\delta_b$. The curves correspond to different top surface disorder cases, changing the disorder strength, $\delta_s$, and concentration, $n_s$, while keeping the bottom surface disorder concentration fixed to $n_b = 1.3$. The inset in the top left corner shows the reciprocal area covered by the surface states in the 9L system. Lines act as guides to the eye.

In support of our observations suggesting a relevant role of the TPSS in the electronic properties, we have carried out two complementary calculations: a ballistic conductance calculation of defect-free FL antimonene contacted on the surface (Figure 4a) and conductivity calculations in the presence of surface disorder (Figure 4b). The results of the conductance (per unit length) are shown in Figure 4a (see Methods for details and Supplementary Figure 11 for several examples of transmission curves). The conductance increases with the number of layers, but at such slow pace that, for 9 layers, it is not even twice that of 3 layers (where only the Dirac cone contributes to the conductance) and visibly saturates. From these calculations, however, we cannot fully conclude whether the slow conductance increase is due to the full development of the TPSS in the center of the BZ or to the increasing presence of bulk states (for instance those at M). In order to assess this issue, we show in Figure 4b the (diffusive) Kubo conductivity of our thicker 9L system in the presence of disorder on both top and bottom surfaces (see Methods for details). We can calculate the conductivity from the TPSS alone or from all the states in our multilayers, surface and bulk.

Here we plot the ratio between the TPSS conductivity of the top surface (actually considering only surface states, namely, excluding the halves of the hole pockets which correspond to bulk states), and the total conductivity (top TPSS + bulk states) for different disorder realizations as a function of the disorder strength affecting the bottom surface. In general, the TPSS contribution represents a significant fraction over the total conductivity (around 50 %), despite introducing "only surface disorder". In fact, increasing the strength of disorder on the bottom layer with respect to the top surface, the relative contribution of the surface conductivity with respect to the total one increases up to 70 % (in our parameter range). This behavior is expected, as bulk states feel the disorder on the bottom layer more effectively, and it is practically (experimentally) relevant since, while the top surface can be treated or cleaned, the bottom one always remains affected by the substrate. This behavior is shared by all disorder cases up to a point where the bottom disorder gets strong enough as to also affect the top surface states in the hole pockets, which cannot be dissociated from the bulk states, reducing the TPSS conductivity. For thicker flakes, this turning point is likely to shift to larger bottom disorder strengths. The overall picture emerging from these two different calculations implies that the electrical transport is amply dominated by the TPSS, as suggested by our experimental results.

To conclude, we show the electrical characterization of mechanically exfoliated FL antimonene flakes within the ~ 2-3 to 100 layers regime, where topologically protected surface states might be playing a key role. We detect a strongly quenched photoemission signal at the $E_F$ for the thinnest flakes and a mainly constant value as the thickness increases. Our DFT calculations show that a transition from low to high surface density of states close to the Fermi level takes place at about 6 layers. We determine a sheet resistance of ~ 1200 $\Omega$ sq$^{-1}$ above this threshold independent on the sheet thickness. We also estimate the mobility of FL antimonene in ambient conditions, resulting in 150 cm$^2$V$^{-1}$s$^{-1}$. Such a high value, in combination with its stability and simple structure, turns antimonene into a promising candidate for nanoelectronics and optoelectronics applications. Hence, we envision that the richness in properties and phenomena of this material (including optoelectronic and energy-related) will pave the way for further studies on its many opportunities both from fundamental and practical points of view, with a particular focus in topological surface state physics, as for example for fault-tolerant quantum computation or as conducting channels with reduced dissipation in spintronic devices.

## 4. Methods

*4.1. Sample preparation.* We obtained FL antimonene flakes by mechanical exfoliation [1]. We placed a macroscopic freshly cleaved crystal of antimony (Smart Elements) in adhesive tape and after repetitive pealing, we directly transferred FL antimonene flakes to $SiO_2$/Si and highly doped Si substrates. A primary optical microscopy inspection allowed us to locate the thinnest flakes that we later imaged by AFM in contact mode to measure their thickness [41]. We used OMCL-RC800PSA cantilevers from Olympus with a nominal spring constant of 0.39 N m$^{-1}$ and low forces of the order of ~ 1 nN to ensure that the flakes were not deformed by the tip. We used WSxM software (www.wsxm.es) both for the acquisition and processing of the AFM data [42, 43].

*4.2. PEEM/LEEM measurements.* The experiments have been carried out at the PEEM experimental station of the CIRCE beamline at the ALBA Synchrotron [44]. All measurements were done in a low energy and photoemission electron microscope from micrometer-sized FL antimonene crystals. Prior to the measurements, we carried out a careful cleaning process in ultra-high vacuum (UHV), consisting of several annealing cycles at 400 ºC for 5 minutes in a hydrogen atmosphere ($10^{-6}$ mbar). This cleaning protocol ensured the flakes were free of oxide and contamination.

*4.3. KPFM measurements.* We carried out simultaneous dynamic mode AFM for the topography and frequency modulation mode for the KPFM [45] in a single-pass scheme, using metallized AFM tips (Budget Sensors ElectriMulti75-G). We applied an AC bias voltage of amplitude 5 V and frequency 7 kHz to the tip. We performed the KPFM measurements in an inert Ar atmosphere to avoid CPD shielding by the presence of adsorbed water on the surface of the samples [36].

*4.4. C-AFM measurements.* We contacted the FL antimonene flakes deposited on $SiO_2$/Si substrates using gold nanowire electrodes through the SPANC technique [37]. Briefly, we deposited gold nanowires on the substrates with FL antimonene flakes and assembled them into nanoelectrodes by AFM manipulation. In this case, the so-fabricated gold nanoelectrodes were connected to a microscopic gold electrode, we then used a metallized AFM tip (Budget Sensors ElectriMulti75-G) as a second mobile electrode to acquire *IV* curves at different locations. We employed dynamic mode AFM to image the samples, with an amplitude set point of 15 nm (cantilever free amplitude 20 nm). Then, we stopped the tip over the points of interest, and we brought it down into contact. There we acquired several *IV* curves and after this we brought the tip back to dynamic mode AFM. Before and after each set of *IV* curves we checked that the tip had not changed by acquiring *IV* curves on the gold nanoelectrode, ensuring tip stability along the whole set of measurements. To avoid artifacts, we also always carefully check the dependence of the conductance with the applied load, selecting the optimal conditions [46]. This procedure ensures that any possible layer of contamination or oxide on the surface is pierced and the current measured would respond to the intrinsic properties of the material [47]. In addition, we obtain the sheet resistance of the few-layer antimonene flakes from the flakes geometry and the slope of the Resistance *vs.* Lentgh (*RL*) plots, which is independent of the contact resistance, thus obtaining reliable sheet resistance values within the uncertainty in the measured currents. *RL* plots acquired on the same flake but with different AFM tips can present different contact resistances, but they present the same slope (and therefore they lead to the same sheet resistance) (see Supplementary Figure 10).

*4.5. Band structure calculations.* Band structure and Fermi surface contour calculations in Figure 1 were performed using DFT as implemented in the Vienna Ab initio Simulation Package (VASP) code [48, 49]. We employed the local density approximation (LDA) [50] together with the projector-augmented-wave (PAW) [49] method which has been shown to correctly describe Sb films and predicts the bulk lattice constant [21] in excellent agreement with the experimental value of 4.30 Å [51]. Calculations are performed considering SOC and a plane wave basis set was used with a cutoff energy of 300 eV on a 21×21 Monkhorst-Pack [52] k-point mesh. A vacuum region of 16 Å along the z direction was used for each system in order to minimize the interaction between the periodic repetitions of the cells. Full structural relaxations were performed for each of the 2L-7L systems until atomic forces were smaller than 0.001eV/Å [53]. The relaxed structure of the 7L system was used for the 9L system employed for the conductivity calculations (see below). We have taken defects into consideration as a perturbation (to all orders), expanding the eigenstates of the system + defects in a basis of Bloch states near the Fermi energy of the defect-free system (see Ref. [27] for details).

*4.6. Ballistic transport calculations.* We have obtained the room temperature conductance of the pristine multilayers thermal averaging the Landauer expression, $G=e^2/h\ T(\mu)$, where the transmission, *T*, is computed with the non-equilibrium Green's function formalism [54]

$$T(\mu) = \frac{1}{Na}\sum_{k}^{N} Tr[\Gamma_L(k,u)G(k,u)\Gamma_R(k,u)G^+(k,u)]$$

We have found that the most convenient way to contact the surface layer (mimicking the experiment) is, as shown in the insets of Figure 4a, to continue to infinity this layer along the *x* direction, but gated with a potential of 2.0 eV that sets the Fermi level into the valence band where its conductance is maximum. The multilayer is finite in the *x* direction (with the maximum length permitted by our computational limitations) and periodic in the *y* direction with a minimal unit cell of width *a* (as shown in the insets). Thus, the Green's functions, G, and coupling matrices, Γ, depend on *k* in this direction. The summation over *N* values of *k* in the first Brillouin zone divided by the actual width in the perpendicular direction, *Na*, gives the conductance per unit length. The Green's functions and coupling matrices are evaluated from the non-collinear DFT Hamiltonian obtained with the OpenMX code [55] (http://www.openmx-square.org/), computing the whole system shown in the insets in Figure 4a periodically repeated. OpenMX is based on norm-conserving pseudo-potential method with a partial core correction and a linear combinations of pseudo atomic orbitals (LCPAO) as a basis function here specified by Sb7.0-s3p2d2 (meaning that the cutoff radius is 7.0 Bohr and that three primitive orbitals for s components, and two primitive orbitals for each of p and d components are used). The exchange-correlation functional was the spin-polarized GGA-PBE. All calculations were performed until the change in total energy between two successive iteration steps converged to less than $10^{-6}$ Hartree. A cutoff energy of 220 Ry and an 11x11x1 k-grid have been used in all presented results. The unit cell structure was also geometrically optimized. The quasi-Newton eigenvector following method was executed for structural relaxation of all structures until the change in forces between two successive iteration steps was less than $10^{-3}$ Hartree/Bohr. SOC was also employed and van der Waals interaction (vdW) correlation is considered by using the semiempirical dispersion-corrected density functional theory (DFT-D2) force-field approach. We have verified that the obtained band structure is essentially similar to that shown in Figure 1 (see Supplementary Figure 12).

*4.7. Kubo conductivity calculations.* We have chosen a system of 9 antimonene layers for the calculation of the Kubo conductivity. This number guarantees that the bulk pockets are well developed and that the TPSS of opposite surfaces are completely decoupled. We first compute the DFT Kohn-Sham Hamiltonian with the CRYSTAL code [56, 57] to which we have added SOC after self-consistency [58]. The lattice constant was considered (based on our previous work, Ref. [58]) to be a = 4.27 Å and the intra- and interlayer distances h = 1.52 Å and d = 3.68 Å, respectively. For these calculations we utilized a small-core pseudopotential basis set [59] with 23 valence electrons. For better agreement with plane-wave calculations we considered 0.94 scaling factor for the last filled p-orbital. The SOC enhancement factor 65, was also defined based on our previous work where we verified the SOC implementation *vs.* fully relativistic non-colinear VASP calculation (Ref. [58]). LDA exchange [60] and VBH correlation [61] functionals were used on a 32 x 32 Monkhorst-pack k-grid. For convergence, we applied a Fock (Kohn-Sham) matrix mixing of 97% between subsequent SCF cycles and the convergence on total energy was set to $10^{-9}$ Hartree.

We calculate the dc conductivity using finite-size Kubo formalism at room temperature. We have essentially followed the methodology developed in our previous work [27], but here we have fully implemented the exact evaluation of momentum matrix elements, as explained in Ref. [62]. The red bands shown in Figure 1 are the relevant ones in our calculations. The exfoliation procedure guarantees that the internal layers of our flakes are defect-free, but not the surfaces. We also expect the bottom layer to be more influenced by disorder than the top surface due to the substrate effects (substrate charges and intercalated molecules). For simplicity, we assume the disorder not to mix states between bulk pockets or between bulk and surface pockets. This way the total conductivity is simply the sum of the individual pocket conductivities. For these calculations we have used a square sample of 45 nm in size and studied different disorder configurations defined by the strength of the top and bottom surface disorder, $\delta_s$ and $\delta_b$, and the top and bottom surface disorder concentrations, $n_s$ and $n_b$, respectively. In order to achieve statistical precision, we have considered at least 100 configurations of disorder on each surface up to the point that the behaviour of the curves in Figure 4b was not changing anymore. Furthermore, we have smoothed the curves by averaging over the Fermi energy interval containing typically 11 levels.

## 5. Acknowledgments

We acknowledge financial support through the "María de Maeztu" Programme for Units of Excellence in R&D (CEX2018-000805-M), the Spanish MINECO through projects PCI2018-093081, FIS2016-80434-P, PID2019-109539GB-C43, PID2019-106268GB-C31 and -C32,

MAT2016-77608-C3-1-P and -3-P, MAD2D-CM, MAT2013-46753-C2-2-P and MAT2017-85089-C2-1R, the EU Graphene Flagship funding (Graphene Flagship Core3 881603 and JTC2017/2D-Sb&Ge), the EU via the ERC-Synergy Program (Grant ERC-2013-SYG-610256 NANOCOSMOS) and the European Structural Funds via FotoArt-CM project (S2018/NMT-4367), the Fundación Ramón Areces and the Comunidad Autónoma de Madrid through S2018/NMT-4321 (NanomagCOST-CM). J.J.P. acknowledges the computer resources and assistance provided by the Centro de Computación Científica of the Universidad Autónoma de Madrid and the RES. S.P. acknowledges financial support by the VILLUM FONDEN via the Centre of Excellence for Dirac Materials (Grant No. 11744).

# Supplementary Information

**S1. Band structure of multilayer antimonene with adsorbed water**

In order to assess the role of environmental conditions on the antimonene band structure, we have carried out DFT calculations with a few layers of water molecules covering the top surface. Somewhat expectedly, the water molecules physisorb and influence the band structure by slightly removing the degeneracy of the top and bottom TPSS. These calculations were carried out with the QUANTUM ESPRESSO package.[1] We used the exchange-correlation functional in the generalized gradient approximation of Perdew-Burke-Ernzerhof.[2] The electron-ion interactions were described by ultrasoft pseudopotentials.

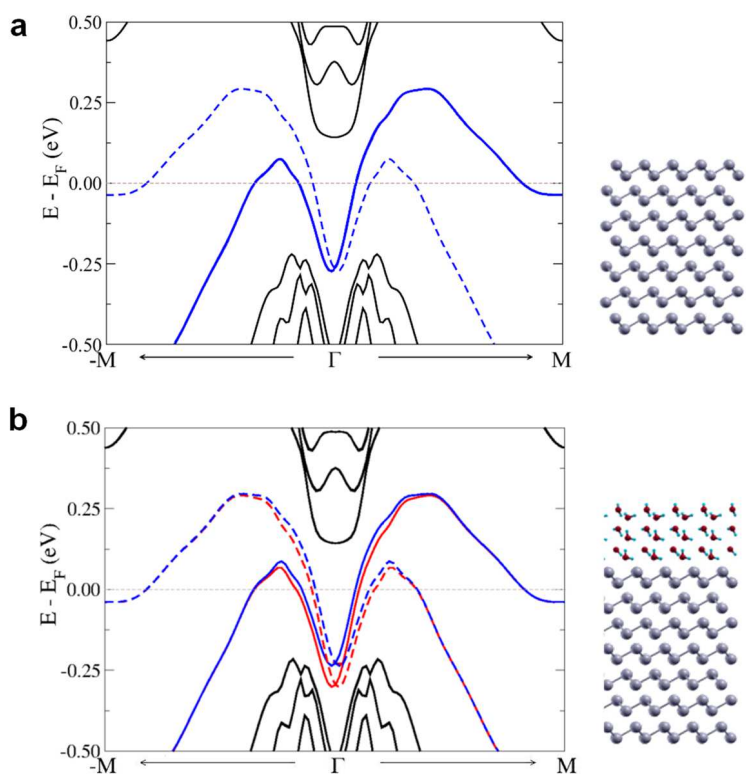

**Supplementary Figure 1. Band structure of 7-layer antimonene. a**, Band structure in vacuum. Solid and dashed lines are degenerate and represent opposite in plane spins near Γ. **b**, Band structure including water molecules on the top surface. Blue and red represent different surfaces when distinguishable. The insets at the right show side views of the atomic configurations used for the calculations.

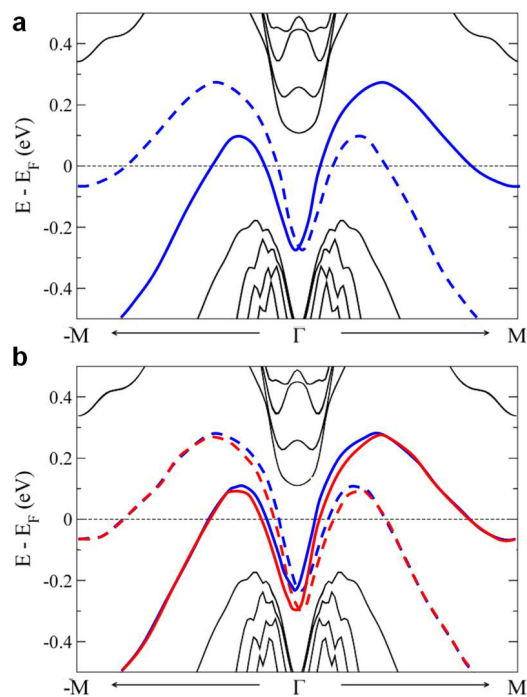

**Supplementary Figure 2. Band structure of 9-layer antimonene. a**, Band structure in vacuum. Solid and dashed lines are degenerate and represent opposite in plane spins near $\Gamma$. **b**, Band structure including water molecules on the top surface. Blue and red represent different surfaces when distinguishable.

Comparison of the exfoliation energies with the observed experimental structure can be informative in some situations[3], although this may not be the case here. We calculated exfoliation energies of antimonene in a previous work[4], but this number would allow to better understand exfoliation in liquid phases, but not so much in a mechanical method, since there is no experimentally controlled quantity to compare with.

**S2. Additional characterization of FL antimonene**

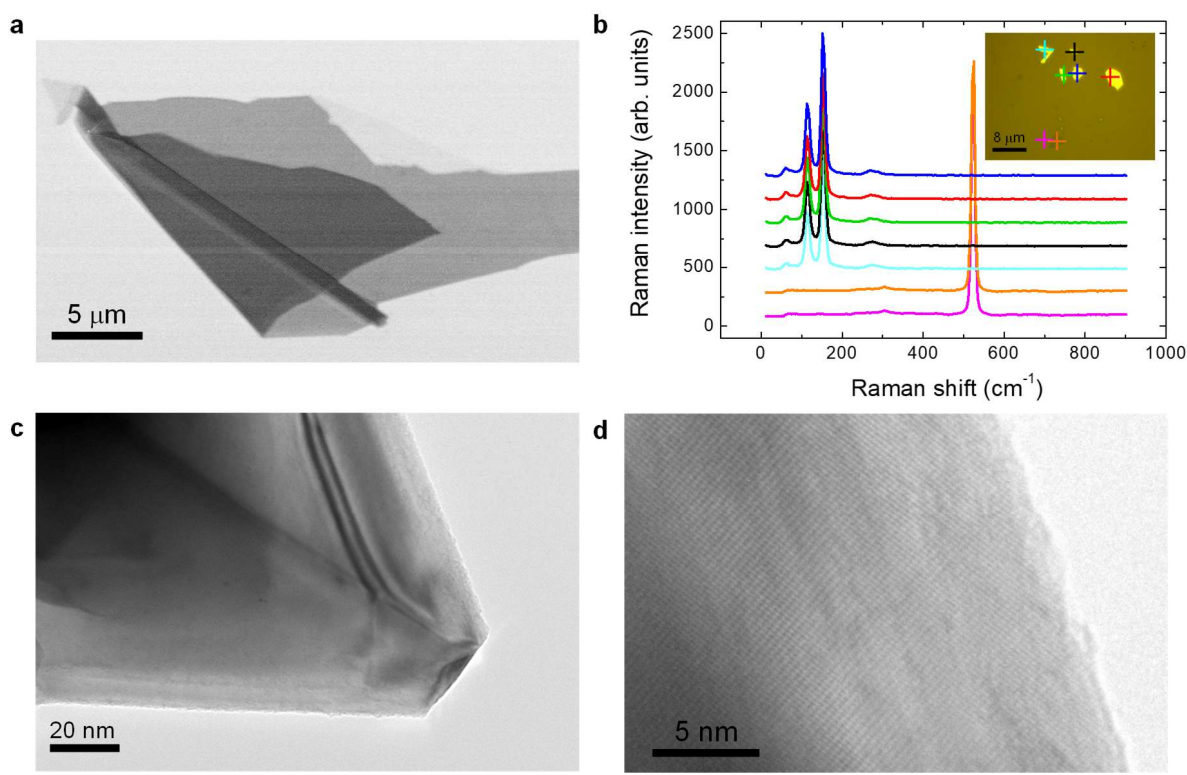

**Supplementary Figure 3. Characterization of FL antimonene flakes with additional techniques a**, Scanning electron microscopy (SEM) image of a part of an FL antimonene flake, in which the contrast is directly related to the thickness (the more contrast the thicker). **b**, Raman spectra from the points corresponding to different FL antimonene flakes in the optical image of the inset. The purple spectrum corresponds to the substrate; the remaining spectra correspond to antimony flakes of different thicknesses. For thin flakes (orange) there is no contrast in the Raman signal [5,6]. **c, d**, Transmission electron microscopy (TEM) images of a thin antimony flake at low (c) and high (d) magnification, showing atomic resolution.

## S3. Low energy and photoemission electron microscopy (LEEM/PEEM), in combination with synchrotron based X-ray photoelectron spectroscopy (XPEEM)

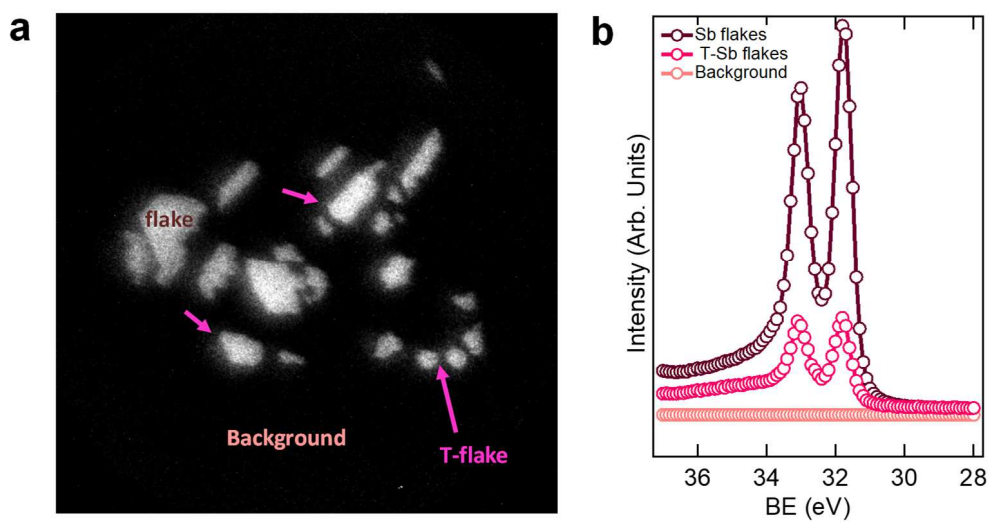

**Supplementary Figure 4. Photoemission electron microscopy combined with X-ray photoelectron spectroscopy. a**, Sb 4d XPEEM image of the same antimony flakes characterized by LEEM and AFM in the main text (see Figure 2) at a photon energy of 97 eV. This chemical mapping, taken at the maximum of the Sb 4d level, points out the chemical composition of the flakes, antimony without traces of oxide or contamination. In the areas corresponding to the thinnest zones of the flakes, 2-5 atomic layers (T-flakes) the intensity is reduced and can be seen as an halo (dark pink arrows). **b**, Sb 4d core level µ-XPS spectra taken in the thickest and the thinnest parts of the flakes and in the background. Despite the intensity in the thicker parts of the flakes (wine spectrum) is higher, as expected in comparison with the thinnest parts (T-flakes, dark pink spectrum), antimony is detected in all of them. In the background (light pink spectrum) no signal of antimony is found.

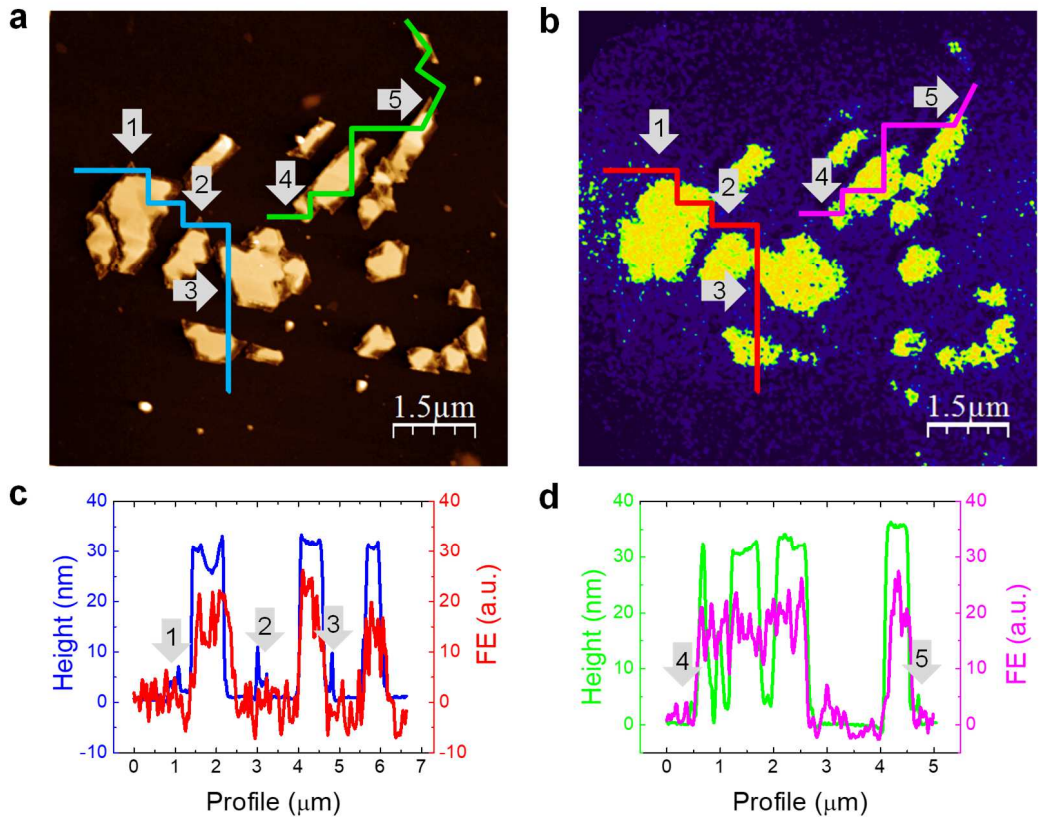

**Supplementary Figure 5. PEEM Fermi Energy level analysis. a**, AFM topographical image of the flakes characterized by PEEM/LEEM. Black to white scale, 40 nm **b**, Image taken at the Fermi Energy level (FE). **c**, **d**, Comparative profiles along the lines in **a**, **b**. Numbered arrows point to FL antimonene terraces below ~ 2.5 nm, where there is no FE signal.

In the PEEM image taken at the Fermi level the contrast can be correlated to the electronic states of the sample. As can be seen, there is not much contrast change across the antimony flakes, indicating same electronic behavior in thicker structures (20-40 nm). However, in a more detailed analysis, in the edges of the flakes where terraces below 2.5 nm can be found (pointed by the arrows in the image) no FE signal is visible. This lack of signal indicates a change in the electronic behavior in comparison with the thicker areas, suggesting a relevant role of TPSS in the few-layer antimonene flakes.

## S4. Kelvin probe force microscopy (KPFM)

We carried out KPFM measurements on flakes with thicknesses ranging from ~ 2 to 9 nm (~ 2-3 to 21 layers).[5] KPFM provides information on the Contact Potential Difference (CPD) variation, related to the work function change.[7] Figure S5 presents the results for two FL antimonene flakes with several terraces of different heights.

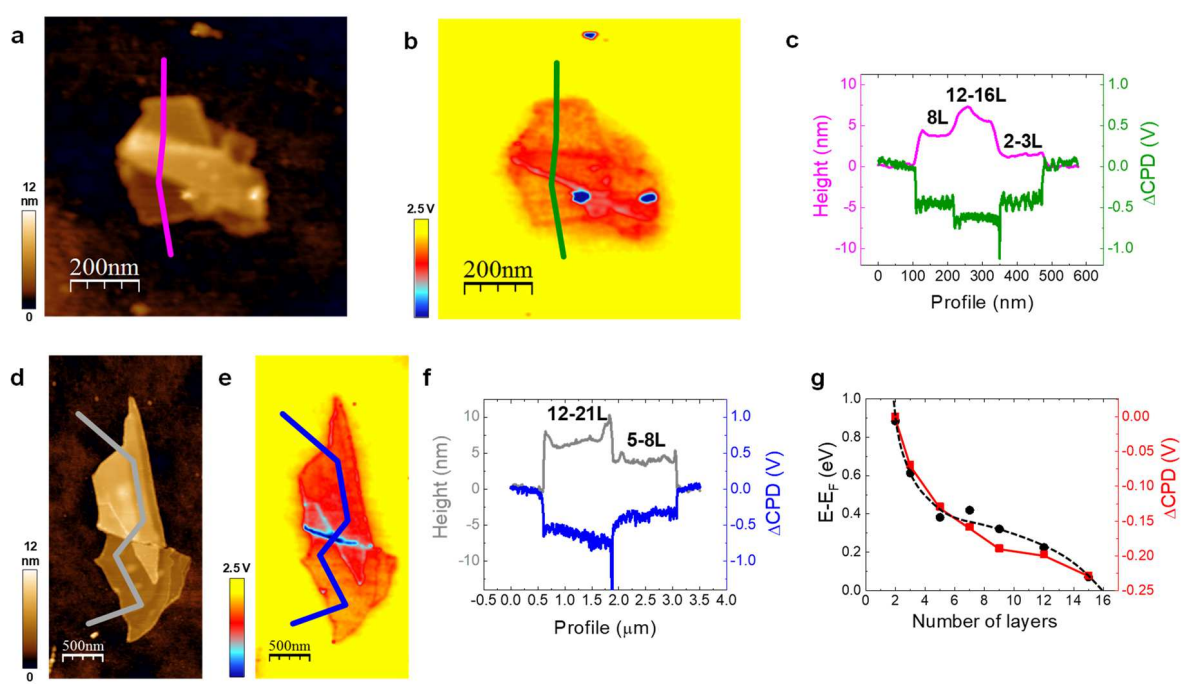

**Supplementary Figure 6. KPFM characterization of FL antimonene flakes. a**, **d**, AFM topographical images of FL antimonene flakes with different thicknesses. **b**, **e**, CPD variation of the corresponding flakes. **c**, **f**, Comparative profiles of the terrace thicknesses and CPD variation along the lines in **a**, **b** and **d**, **e** (the lines in **b** and **e** indicate schematically the path followed to perform the profiles, see Figure S6 for details). **g**, DFT calculations of the energy value of the bottom of the second bulk conduction band (black circles and dotted line) and the CPD variation (red squares and solid line) as a function of numbers of layers. Lines act as guides to the eye.

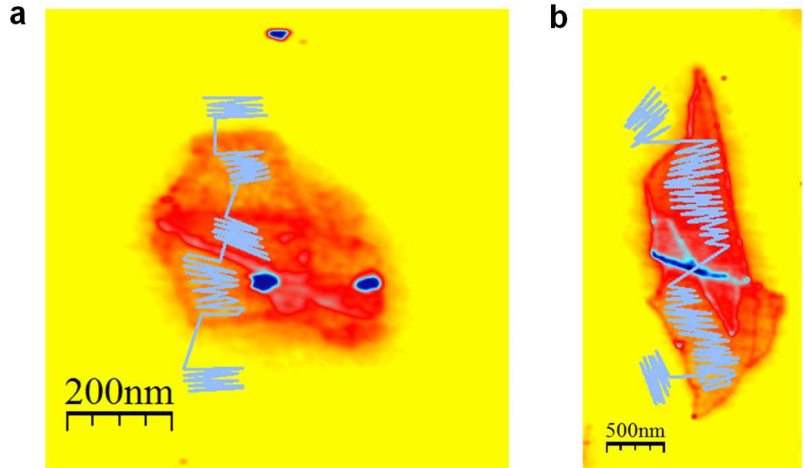

**Supplementary Figure 7. KPFM profiles. a**, **b**, CPD variation of the FL antimonene flakes. The blue lines correspond to the real paths followed to plot the profiles of the variation of the CPD shown in Figure S5. The length for each section of the CPD profiles is proportional to the corresponding section in the topographical profiles. Thus, we could re-scale the CPD profiles to match the total length of the topographical profiles. Following this procedure, the CPD profiles shown better reflect the overall behavior of the flakes compared to having used direct profiles.

For terraces above ~ 8 layers the CPD has appreciably decreased. The variation of the CPD contrast is inversely proportional to the variation of the work function in the sample ($\Delta$CPD = -$\Delta\phi_{sample}$ sample/$e$, where $e$ is the magnitude of the electron charge), determined by the relative position of the Fermi levels, which are sensitive to the surface charge density or the presence of surface dipoles.[7] The measured CPD is higher for terraces below ~ 8 layers and lower for thicker ones. Our DFT calculations in Figure S5g show this behavior, with the CPD actually decreasing with the number of layers, compatible with the KPFM measurements.

## S5. Conductive AFM measurements

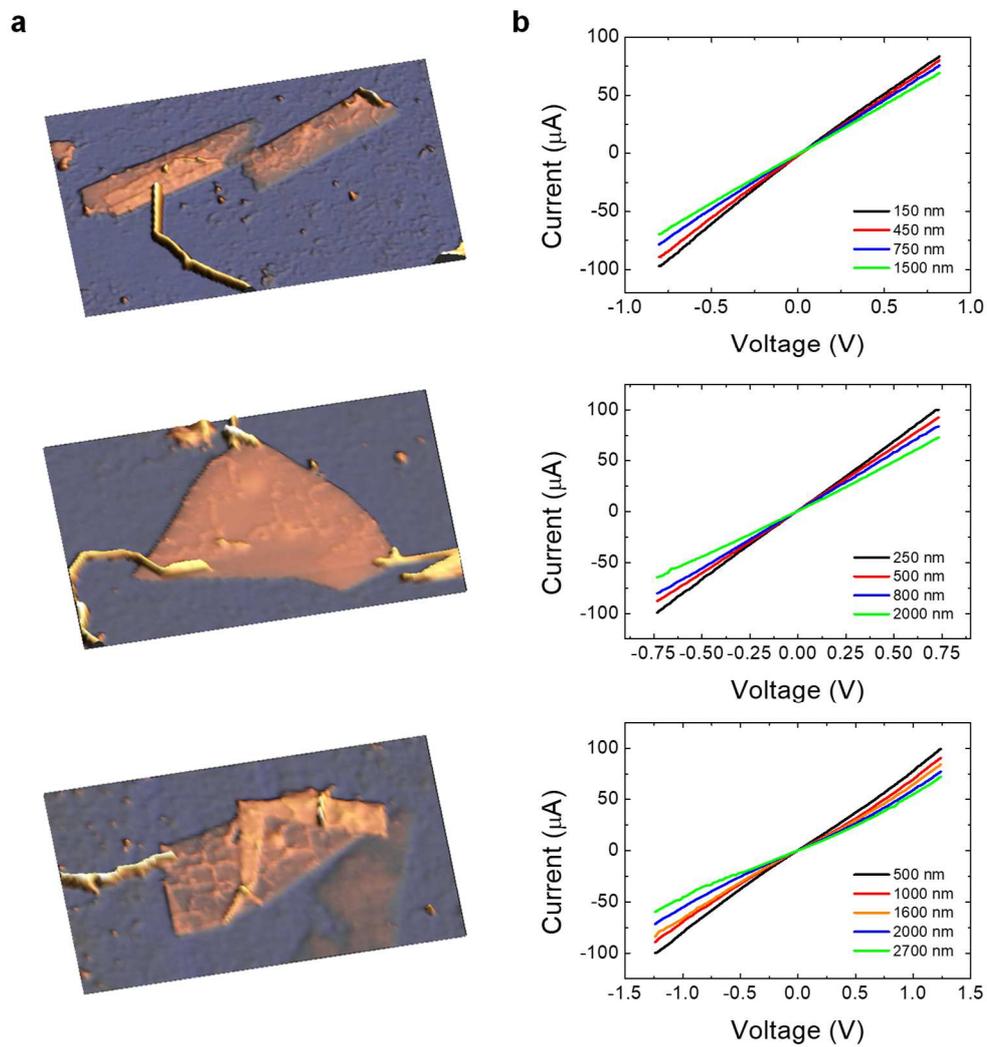

**Supplementary Figure 8.** *IV* **curves at different distances. a**, AFM topographical images of several FL antimonene flakes, similar as in Figure 3 in the main text. **b**, *IV* curves acquired at different distances from the electrode for each of the flakes.

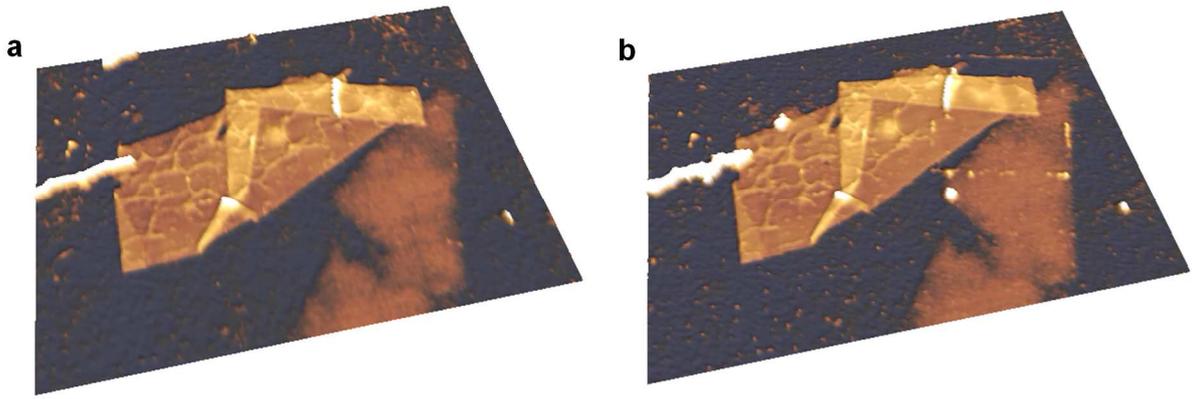

**Supplementary Figure 9. Stability under ambient conditions. a**, AFM topographical images of the same FL antimonene flake after exfoliation and **b**, more than one year later, showing no noticeable degradation.

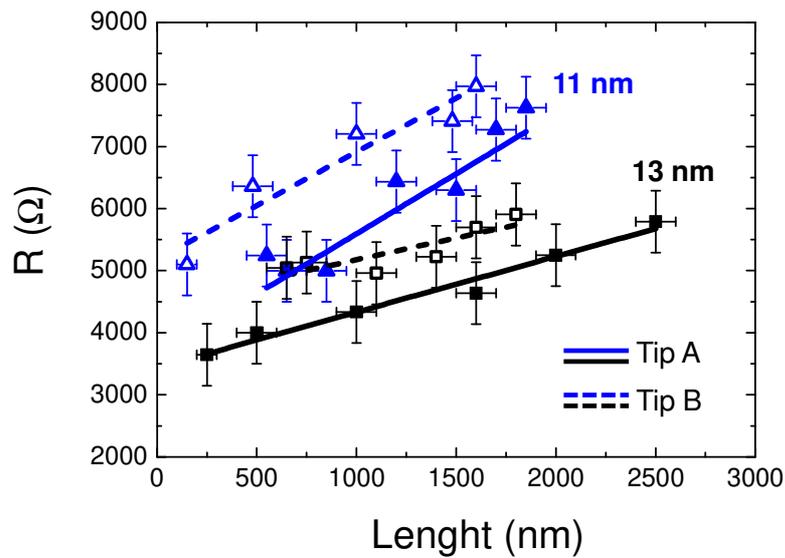

**Supplementary Figure 10. Resistance *vs.* Length (*RL*) plots for 11 and 13 nm thicknesses acquired with two different conductive AFM tips as indicated in the graph.** Lines are the best linear fits. As can be seen, each thickness presents the same slope in the *RLs* independent of the tip, with only an offset difference coming from different contact resistances.

## S6. Ballistic conductance calculations

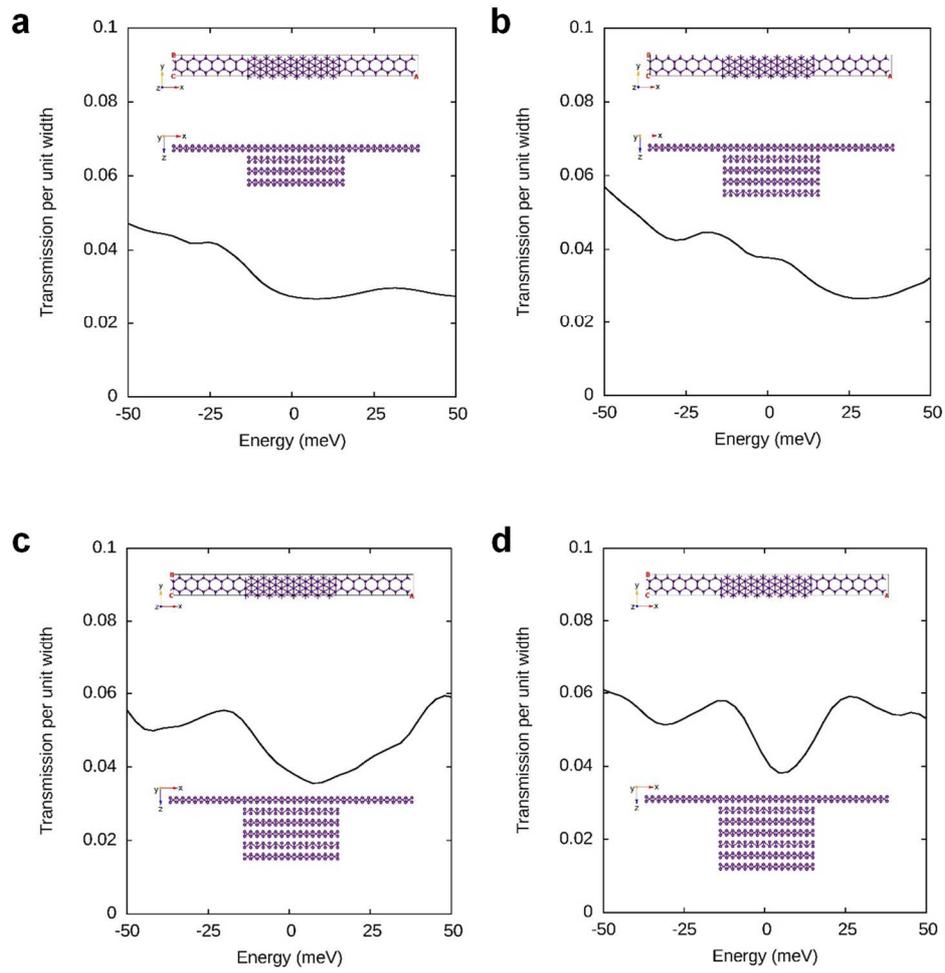

**Supplementary Figure 11. Transmission for contacted multilayers.** Several examples of transmission curves obtained for different number of layers (4, 5, 6, and 7 layers for **a**, **b**, **c** and **d** respectively).

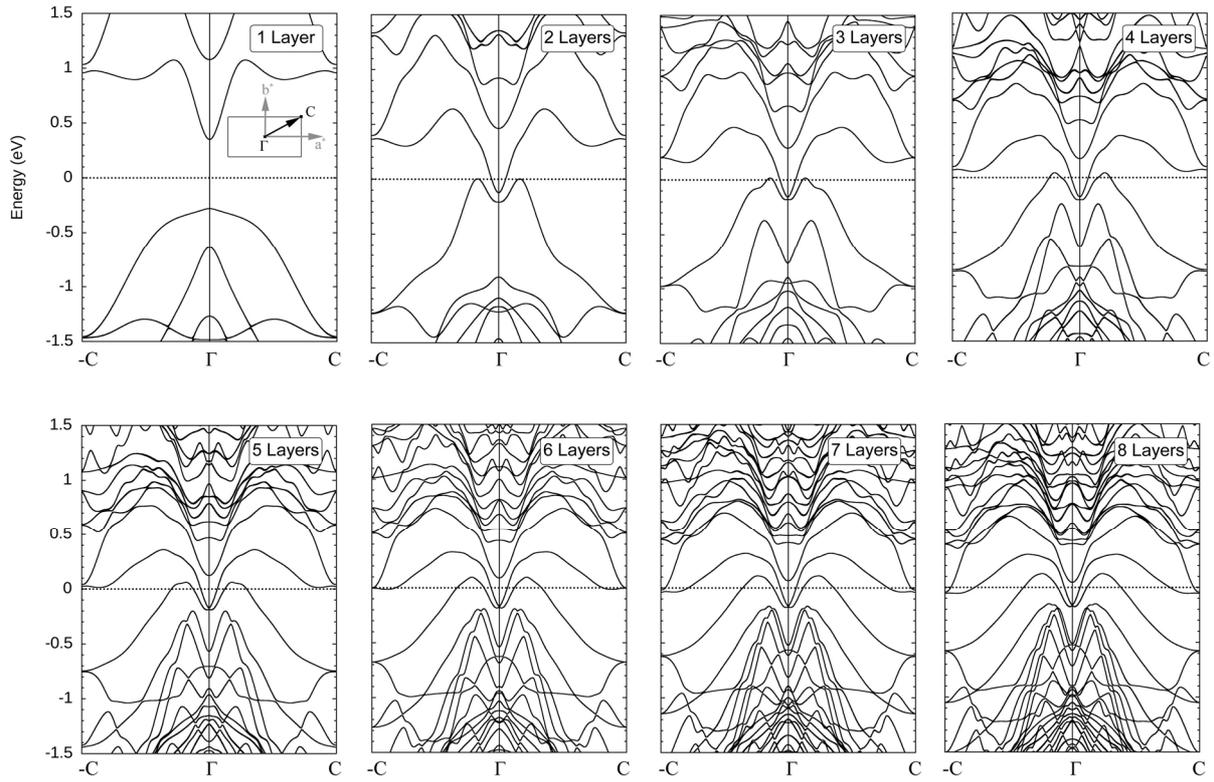

**Supplementary Figure 12. Band structure of FL antimonene based on the OpenMX code.** An almost perfect match with the VASP bands in Figure 1 in the main text can be observed.